\documentclass[aps,prl,twocolumn,noshowpacs,superscriptaddress,groupedaddress]{revtex4}  
\usepackage{graphicx}  
\usepackage{dcolumn}   
\usepackage{bm}        
\usepackage{amssymb}   

\usepackage[T2A]{fontenc}

\usepackage[utf8]{inputenc}

\usepackage[english,russian]{babel}

\usepackage{amsmath}

\usepackage{amsfonts}

\usepackage{textcomp}

\usepackage{bm}

\usepackage{color}

\usepackage{verbatim}

\usepackage{subfig}

\begin{document}

\title{On Magnetostatics of Chiral Media}

\author{Z.V. Khaidukov}
\affiliation{ITEP, B. Cheremushkinskaya 25, Moscow, 117218 Russia}
\author{V.P. Kirilin}
\affiliation{ITEP, B. Cheremushkinskaya 25, Moscow, 117218 Russia}
\author{A.V. Sadofyev}
\affiliation{ITEP, B. Cheremushkinskaya 25, Moscow, 117218 Russia}
\author{V.I. Zakharov}
\affiliation{ITEP, B. Cheremushkinskaya 25, Moscow, 117218 Russia}
\affiliation{Max-Planck Institut f\"{u}r Physik, 80805 M\"{u}nchen, Germany;}
\affiliation{Moscow Inst Phys \& Technol, Dolgoprudny, Moscow Region, 141700 Russia.}

\begin{abstract}
We consider  magnetostatics of chiral media with a non-vanishing chiral
chemical potential $\mu_5\neq 0$. The chiral anomaly is known to have macroscopic 
manifestations which go beyond the standard classical electrodynamics and we introduce
an effective action which accounts for the effect of the anomaly.  A new piece in the effective action takes the form of a 
topological three-dimensional
photon mass. The 
topological mass of the photon turns to be imaginary and signals instability. A stable state
corresponds to a solution of the Beltrami equations.
We demonstrate also that the interaction between two current loops reduces to the
linking number of the loops. As for the chiral magnetic effect it seems to disappear
in the far infrared.  
\end{abstract}

\maketitle

\section{Introduction}

Electrodynamics of chiral media attracted a lot of attention recently. One expects that
in such media the triangle chiral anomaly, which is a pure quantum phenomenon, has
macroscopic manifestations,
for a review and further references see \cite{review}.
 In particular,
an external magnetic field ${\bf B}$ induces an electric current proportional to $\mu_5$ \cite{kharzeev}:
\begin{eqnarray}
\label{cme}
J_i = \sigma B_i
\end{eqnarray}
where  $\sigma$  can be referred to as the chiral conductivity.
The chiral  conductivity can readily be evaluated on one -loop level, $\sigma
= (e^2\mu_5)/(2\pi^2)$ . Moreover, there exist beautiful theorems on non-renormalizbility
of $\sigma$ rooted in the non-renormalizability of the triangle anomaly.
As first argued in \cite{surowka}  the non-renormalizabilty of $\sigma$ persists
even in the hydrodynamic approximation. 

Another remarkable feature of the chiral magnetic effect (\ref{cme})
is that the current is non-dissipative, as a consequence of the time-reversal
invariance of strong interactions \cite{yee}. In this respect, the chiral conductivity is
similar to the Hall conductivity (for a review and references see, e.g., \cite{hall}).
The non-dissipativity of the current (\ref{cme}) might become a basis for new
technologies \cite{technology}. Moreover, there is no apparent temperature dependence
of the chiral conductivity.

Non-dissipativity is usually related to topological nature of the corresponding effects,
see, e.g., \cite{hall}. It would be important therefore to clarify the topological origin,
if any, of the chiral magnetic effect (CME), especially in view of the absence of any apparent
temperature depedence of the chiral conductivity. At vanishing temperature
there exist derivations of the CME in terms of the Berry phase
in the momentum space \cite{yamamoto}
 which shed some light on the topology related to the chiral magnetic effect.

It is crucial  to emphasize that the chiral magnetic effect is a feature of the equilibrium
physics and, in field-theoretic language, represents a static effect.
Namely,  the chiral conductivity,
to the linear order in the external field, is given by the
antisymmetric part of the current-current correlator at frequency identical zero and
3d momentum tending to zero, $\vec{k}\rightarrow0$:
\begin{eqnarray}
\label{sigmaCME}
\sigma = \lim_{k_n\to 0} \epsilon_{ijn} \frac{i}{2 k_n} \Pi^{ij},
\end{eqnarray}
where $\Pi_{ij} = \langle J_i,J_j\rangle$.
Eq. (\ref{sigmaCME}) is to be contrasted with  the standard Kubo formulae
which assume another limiting procedure,
 $\vec{k} = 0, \omega \rightarrow 0$, and
define transport coefficients dynamical in origin.  The chiral conductivity
is actually static, suggesting the recently coined name "thermodynamic
susceptibility'' to describe it \cite{yamamoto,jensen}.

In this note we concentrate, therefore, on the long-range forces in the static limit.  
The point is that the chiral magnetic effect
(\ref{cme}) modifies in fact the Maxwell electrodynamics. Namely, one should take into account
the backreaction of the medium, or the magnetic field generated by the current (\ref{cme}).
The modification turns drastic at large distances of order $r\sim  (e^2\mu_5)^{-1}$.

We perform the calculations using an effective action which takes
 into account the triangle anomaly: 
\begin{equation}\label{1loop}
\delta S_{eff}~=~\frac{1}{2}\sigma A_i\partial_jA_n\epsilon_{ijn}~,
\end{equation}
where $A_i$ is the vector potential of electromagnetic field.
Effective action has been widely used, in particular, in thermodynamic
studies of the chiral effects, see, e.g., \cite{minwalla}.
 Similar expansions are common in theory of the Hall effects.
Note, however, that we do not have a gap and, therefore, the validity of the effective-action
approximation, or expansion in momenta is a subtle issue, to our mind. 
A closer look into the dynamics is needed. Here we note only that
there exist schemes where the value of the chiral magnetic effect is related
to the details of ultraviolet renormalization, or polynomials in the correlator of the currents
(\ref{sigmaCME}), see, e.g. \cite{landsteiner}.  
Within such schemes the validity of the effective-action
approximation is granted.

  We perform calculations 
assuming the validity of (\ref{1loop}). Amusingly enough, in this approximation 
the magnetostatics of the four-dimensional (4d)  electrodynamics of chiral media
is described by the Euclidean version of the 3d electrodynamics with  the so called
topological photon mass considered in many papers,
see in particular \cite{jackiw,hill,kogan}.  From our perspective, 
it is the topological aspects and
non-renormalization theorems of the 3d theory which are most relevant.
In particular, it was emphasized in \cite{kogan} that the 3d topological photon mass
corresponds to an instantaneous interaction, and this observation  adds insight to
understanding the non-dissipativity of the chiral magnetic effect. The non-renormalizability
of the 3d topological mass \cite{hill} has already been exploited
in  the  evaluation of the  chiral vortical effect in the 4d case \cite{golkar}. 
Below, we will consider another application
of the technique of the Ref. \cite{hill}. 

The most specific feature of the case considered is that the
3d topological photon mass turns to be imaginary, thus signaling instability
of the external magnetic field in the chiral media.
In chiral media, magnetic fields are rearranged to a kind of a
self-consistent distribution with nonzero field helicity $\mathcal{H}$:
\begin{eqnarray}
\mathcal{H}=\int \vec{A}\cdot \vec{B} d^3x,
\end{eqnarray}
which should be included into the helicity conservation law. 
With this rearrangement, the chiral charge $Q_5$ of the constituents decreases,  
see also \cite{yamamoto1,zamaklar}.

Furthemore, we will also demonstrate that the term
$i \sigma \epsilon_{ijk}k^n$ in the one-loop current-current correlator induces
a similar term in the {\it exact} propagator for the abelian gauge field,
\begin{eqnarray}
\label{polenew}
\delta D_{ij} (\omega = 0, \vec{k}) \rightarrow -
\frac{i \epsilon_{ijl} k^l}{\sigma k^2},~~~\vec{k}\rightarrow0~.
\end{eqnarray}
The term (\ref{polenew})
results in a topological type interaction of static current loops ($k\ll \sigma$), 
proportional to their linking number. 

Moreover, it is well known that the exact photon propagator is related to the
exact current-current correlator in the following way:
\begin{eqnarray}
\label{corr}
D_{ij} = D_{(0)ij} + D_{(0)ik} \Pi_{kl} D_{(0)lj},
\end{eqnarray}
which implies that
\begin{eqnarray}
\label{CME}
\Pi^{A}_{ij} =  -i \frac{k^2}{\sigma} \epsilon_{ijl}k^l + O(k^4),
\end{eqnarray}
where $A$ stands for the antisymmetric part.  
Note the appearance of $\sigma$ in the denominator. 
According to (\ref{sigmaCME}), this result implies that in the strict $\vec{k} \rightarrow 0$
limit the CME current actually vanishes.

\section{Propagator}

Let us start by setting up notations.
Here and below we  use the metric
$g_{\mu\nu} = diag(-,+,+,+)$. The theory we will consider is QED with
chiral fermions at a non-zero value of $\mu_5$. The corresponding action is given by:
\begin{eqnarray}
S = \int dt d^3x
[\bar{\psi} (i \gamma^{\mu} D_{\mu}+ \mu_{5} \gamma^{0} \gamma^{5})\psi - \frac{1}{4}F_{\mu\nu}F^{\mu\nu}].
\end{eqnarray}
Here $D_{\mu} = \partial_{\mu} - ieA_{\mu}$.
As is already mentioned above, all the quantities we will calculate are taken at $\omega=0$. 
This puts us into an effectively 3d situation. 
Also, at $\omega=0$ all Minkowski-space two-point functions 
(of two gauge fields or two currents) coincide with the Euclidean ones.

Consider now the dressed photon propagator. 
Rotational and gauge invariances fix its form as:
\begin{eqnarray}
\label{prop}
D_{ij}(0, \vec{k})=D_{S}(k)(\delta_{ij}-\hat{k}_{i}\hat{k}_{j})+D_{A}(k)i\epsilon_{ijl}\hat{k}^l
+ a\frac{\hat{k}_{i}\hat{k}_{j}}{k^2},
\end{eqnarray}
 where $\hat{k}$ is the unit vector along the 3d momentum 
 and the last term is the gauge fixing.
Similarly, the inverse photon propagator obeys
\begin{eqnarray}
\label{propinv}
D^{-1}_{ij}(0, \vec{k})= P_{S}(k)(\delta_{ij} -  \hat{k}_{i}\hat{k}_{j})+P_{A}(k)i\epsilon_{ijl}\hat{k}^l
 + \frac{k^2}{a}\hat{k}_{i}\hat{k}_{j}.
\end{eqnarray}

It is straightforward to show that
\begin{eqnarray}
\label{connS}
D_{S}= \frac{P_{S}}{P_{S}^2-P_{A}^2}.\\
\label{connA}
D_{A}=\frac{P_{A}}{P_{A}^2-P_{S}^2}.
\end{eqnarray}
Moreover, it is well known that the difference between inverses of the
exact and bare photon propagators is given by the sum of 1-particle-irreducible (1PI)
graphs of the photon self-energy,
\begin{eqnarray}
\label{P}
D^{-1}_{ij} = D^{-1}_{(0)ij} - P_{ij}~.
\end{eqnarray}

We shall use this relation to find the infrared limit of the functions $\Pi_S$ and $\Pi_A$.
At one loop $P_{ij} = \Pi_{ij}$, and hence
\begin{eqnarray}
\label{PA}
P_A = - \sigma k+O(k^2),
\end{eqnarray}
as follows from (\ref{CME}). As for the symmetric part,
\begin{eqnarray}
\label{PS}
P_S = O(k^2),
\end{eqnarray}
to maintain the gauge and Lorentz invariances.

If we keep only linear  terms  in the functions $P_A, P_S$,
the propagator  (\ref{prop}) takes a particularly simple form, corresponding to the summation
of the bubble-type diagrams:  
\begin{eqnarray}
\label{prop1loop}
D_{ij}(0, {k_i})= \frac{1}{k^2-\sigma^2}(\delta_{ij} -  \frac{k_{i}k_{j}}{k^2}) -
\frac{i \sigma \epsilon_{ijl}k^l}{k^2(k^2-\sigma^2)} + a\frac{k_{i}k_{j}}{k^4}.
\end{eqnarray}
This means that the photon has acquired an imaginary mass, $m = i\sigma$ 
which results in an instability which will be discussed later.
Moreover, a novel pole at $k^2=0$ appears in (\ref{prop1loop}). As we show in the next section,
it is responsible for another type of topological interaction in the infrared limit.

If we include terms beyond linear in (\ref{PA}) and (\ref{PS})
and all the diagrams of  higher order in $e^2$, the propagator (\ref{prop1loop})
would be modified.  We will now demonstrate that in spite of that the relation (\ref{polenew})
is exact and unaffected by the interactions.  It is sufficient 
to this end to prove that (\ref{PA})
is actually valid to all orders in $e^2$. The proof is along the lines of
the Coleman-Hill theorem\cite{hill,golkar}. The main idea is to construct all
the diagrams contributing to $P_{ij}$ using the n-photon effective vertices
\begin{eqnarray}
\label{Gamma}
\Gamma^{(n)}_{\mu_1...\mu_n}(k_{(1)}...,k_{(n)}).
\end{eqnarray}
These vertices determine the dressed propagator.
 Indeed, if we cut all internal photon lines then a generic diagram
would be obtained by multiplying various $\Gamma's$ by the corresponding
propagators, and performing loop integrations when necessary.

Next, we proceed by noting that it is sufficient to study
$\Gamma^{(n)}$ in the Euclidean space (our external momenta
are Euclidean by construction, and we can always Wick rotate the internal ones).
Then we can claim the analyticity of $\Gamma^{(n)}$ as a function of its arguments.
The role of an infrared regulator protecting us from the divergences which might stem
 from vanishing fermion mass is played by $\mu_5$. In
complete analogy with \cite{hill} we conclude then
\begin{eqnarray}
\label{GammaO}
\Gamma^{(n)}_{\mu_1...\mu_n}(k_{(1)},...,k_{(n)}) = O(k_{(1)}...k_{(n)}), ~~n > 2.
\end{eqnarray} 
We may now use this observation for a few purposes. First,
if we start contracting the internal photon lines we would find out that
for every factor  ${k_{(i)}^{-2}}$ associated with 
a propagator, two factors of $k_{(i)}$ come from the
$\Gamma's$ to which the photon is attached. This protects us from additional infrared singularities.

Second, and most important, we immediately see that all graphs which involve
only $\Gamma^{(n)}$ with $n > 2$ can only contribute $O(k^2)$ terms to the $P_{ij}$. 
Indeed, if both external lines, carrying momenta
$k$ and $-k$ are attached to a single fermion loop, then the graph is $O(k^2)$
by the formula in the preceding paragraph.
If the lines are attached to different loops then 
a factor $O(k)$ is associated with each of them.
The internal photon propagators do not spoil this due to the arguments in the 
preceding paragraph.
The last subtlety we need to address is the reason, why we need to consider
$\Gamma^{(n)}$ only with $n > 2$. By definition, $P_{ij}$ only involves 1PI diagrams,
 which cannot include an isolated $\Gamma^{(2)}(k, -k)$, where $k$
is the external momenta, because that would mean we could cut one of the photon
lines coming out of this $\Gamma^{(2)}$ to obtain a reducible diagram.
Now all non-isolated $\Gamma^{(2)}$'s,
which carry the loop momentum $k_{(i)}$
would only emerge as dressing the bare propagator connecting other
$\Gamma^{(n)}, n > 2$  effective vertices.
Upon summing up all the diagrams we would arrive
to the conclusion that the worst singularity it can contribute is just
$k^{-2}_{(i)}$
which would cancel with the corresponding factors of $k_{(i)}$ coming from
effective vertices (as stated in the paragraph above).

This concludes our proof that $\Pi_A = \sigma k+O(k^2)$.
At one loop this formula holds, and all higher order corrections
behave as $O(k^2)$. Similarly, $\Pi_S = O(k^2)$.
This is a generalization of the well-known fact of
the non-renormalization of the topological mass term in
2+1 dimensions \cite{hill}.  Substituting
these expansions into (\ref{connA}) we get $D_A = (\sigma k)^{-1} + O(1)$.
 Thus we have obtained a pole (\ref{polenew}) just as in \cite{kogan}.
We will proceed to study the implications of this statement in the next section.

\section{Long-range force}

As shown in \cite{kogan}, the existence of the pole (\ref{polenew}) induces a
new interaction term for two external currents, which has the form of the linking number.
 To see this, we first study the long distance limit of the field produced by external sources
$J_i (\omega=0, \vec{k})$, which we take to be static and purely  three-dimensional ($J_0 = 0$)
from the start. Using the propagator (\ref{polenew}) we get
\begin{eqnarray}
\label{inducedfield}
A_i(\vec{k})=-i\frac{\epsilon_{ijk}k^{k}}{\sigma k^2}J_j(\vec{k}) + ...
\end{eqnarray}
thus if two external currents are separated by a large distance 
their interaction potential includes contribution of the form
\begin{eqnarray}
\label{potential}
\delta V=\frac{2}{\sigma}\int d^3x d^3y \ \epsilon_{ijk}\frac{(x-y)^i}{4 \pi |x-y|^3}J^j(x)J'^k(y).
\end{eqnarray}

Suppose now that these currents have a form $J_{i}(x) =
 I \int d\tau \delta^{3}(\vec{x}-\vec{x}(\tau)) \dot{x}_{i}(\tau)$,
essentially representing an infinitely thin static loop with the
total current (total charge passing through the cross section
per unit time) equal to $I$. If one substitutes this into (\ref{potential}) one gets:
\begin{eqnarray}
\label{linking}
V=\frac{2II'}{\sigma} \int_{C} \!\!\int_{C'} dx^i dy^j \epsilon_{ijk} \frac{(x-y)^{k}}{4 \pi |x-y|^3}
\end{eqnarray}
The double integral in the expression above is proportional to the
Gauss linking number of the two current loops. Thus, to reiterate,
we have obtained that static current loops interact topologically,
that is insensitive to the distance between the loops.
The topological term in (\ref{prop1loop}) is exact in the infrared limit.
 For currents varying faster
this interaction does  get considerable corrections. Similarly, exact values of terms
having a pole at  $k^2\sim \sigma^2$   depend on details
of interaction. It makes one to suggest that if theory is expanded around the proper vacuum 
the topological  term would take the same form
while pole terms will be changed to be regular.

We have shown that interaction between two currents at large distances
has the topological origin and does not depend on distances but depends
on the  topology of space distribution of the currents. It should be mentioned that
(\ref{potential}) is not a 4d topological number 
and could be changed by dynamics of the system.

\section{On the consistency of the static approximation}
As we already noticed,
  the results obtained for the photon propagator can be
also neatly expressed in terms of an effective action.
If we put all our fields to be static from the beginning,
then the low-energy effective action obtained 
\footnote{Derivation of (\ref{action}) is actually worth of scrutinizing further.
The point is that the infrared cut-of in integration over fermions 
is provided, in the absence of fermion mass, by $\mu_5\neq 0$. However,
if one expands in $\mu_5$, then terms of second and higher orders are
infrared unstable and, for example, in the hydrodynamic approximation 
the cut off is
provided by energy density and pressure, see \cite{shevchenko}.
Thus, there could exist a ``hidden" infrared sensitivity of the chiral magnetic
effect and of the action (\ref{action}), for further references see
\cite{vz}.}
after integrating the fermions out would look like:
\begin{eqnarray}\label{action}
S = \int d^3x [ - \frac{1}{4}F_{ij}F^{ij} + \frac{\sigma}{2} \epsilon^{ijk} A_{i} \partial_{j} A_k].
\end{eqnarray}
This is the Euclidean version of the well-known 3d Maxwell-Chern-Simons theory.
A novel point, however, is that the mass is imaginary. 
As a result, the propagator (\ref{prop1loop})
has a pole at $k^2=\sigma^2$, i.e. in the physical region.

At least naively, the presence of the pole at $k^2=\sigma^2$ indicates that
in the true vacuum there are static waves with fixed $k^2$. Although this result
might look bizarre at first sight, fixed values of the momentum is indeed a feature
of the solutions to the Beltrami equation: 
\begin{equation}\label{beltrami}
{\bf curl} ~{\bf B}~=~\sigma {\bf B},
\end{equation}
which can be derived from the effective action (\ref{action}).
Moreover, the simplest solution, the so called Arnold-Beltrami-Childress flow \cite{beltrami},
is characterized by three vectors $\vec{k}_{1,2,3}$ 
which form a left- or right-handed vector basis.
Emergence of such a handedness in the true vacuum is indicated by presence of the pole
at $k^2=\sigma^2$ (see Eq. (\ref{prop1loop}))
in front of the structure $\epsilon_{ijl}k^l$.

The physical picture corresponding to the magnetic field rearrangement is as follows. Chiral particles
move chaotically without magnetic field. With field turning on, particles momenta become ordered 
along the field
and helicity changes macroscopically. The current of chiral particles results, in turn, in a
 magnetic
field distribution with macroscopic nonzero helicity. Finally,  a self-consistent distribution of
the currents and magnetic fields is set in. 
Through the process of evolution to the equilibrium distribution
chiral charge of the constituents decreases while helicity of magnetic field configuration is increasing.

\section{Conclusions}
In this note we have demonstrated emergence of a novel
term in the interaction potential of  static current loops in
presence of axial chemical potential. It has the form of
the Gauss linking number and thus is of the topological origin.
 As all topological phenomena it is essentially long-range, and
we suppose that it may be responsible for the non-dissipativity of the chiral magnetic effect.

We wish to emphasize that essentially we have found an infrared-exact result in the theory. 
In fact,
it is not surprising that it stems from topology, since there are known examples of such an interplay.
One can expect, in general, that there are other quantities which would be fixed in the
same manner, at least in the infrared.

In this sense, the effective action relevant to the magnetostatics belongs to the same sequence as 
effective action relevant to the Hall effect and this observation supports the idea that
the chiral magnetic effect is topological in origin. An important difference, however, is that
the effective action indicates to the instability of any external magnetic field in chiral media.
Furthermore, it is natural to assert that a distribution of magnetic fields,
of the magnitude inverse proportional to electric charge, 
is generated spontaneously
in chiral media with $\mu_5\neq 0$. 
The stable state could either be a smooth macroscopic solution
to the Beltrami equation or, even, a kind of a turbulent or chaotic state, 
as is typical for  the solutions of the Beltrami   equations, see, e.g., \cite{beltrami}. 
\section{Acknowledgments}
The authors are grateful to M.I. Polikarpov and O.V. Teryaev for useful discussions. They also acknowledge helpful remarks of V.A. Rubakov. The work of AVS and VPK has been partly supported by FAIR program for PhD students.

\end{document}